\documentclass[aps,prl,twocolumn,superscriptaddress,longbibliography]{revtex4-2}

\usepackage{bbm}
\usepackage{mathrsfs}
\usepackage{amsmath}
\usepackage{amsfonts}
\usepackage[colorlinks=true,linkcolor=blue,urlcolor=blue,citecolor=blue,anchorcolor=blue]{hyperref}
\usepackage{graphicx,epstopdf}
\usepackage{subfigure}

\usepackage{dcolumn}
\usepackage{bm}
\usepackage{color}
\usepackage{natbib}
\usepackage{amssymb}
\usepackage{xcolor}
\usepackage{braket}
\usepackage[normalem]{ulem}
\usepackage{float}
\usepackage{lipsum}
\usepackage{pifont}
\usepackage[T1]{fontenc}  
\usepackage[utf8]{inputenc}

\definecolor{newtxtcolor1}{rgb}{0.8, 0, 0.2}

\begin{document}

\title{From Spectral Singularities to Multipartite Entanglement Scaling at Higher-Order Exceptional Points}   

\author{Chunlai Yang}
\affiliation{State Key Laboratory of Artificial Microstructure and Mesoscopic Physics, School of Physics, \\Frontiers Science Center for Nano-optoelectronics $\&$ Collaborative Innovation Center of Quantum Matter, Peking University, Beijing 100871, China}
\affiliation{Hefei National Laboratory, Hefei 230088, China}

\author{Shuheng Liu}
\affiliation{State Key Laboratory of Artificial Microstructure and Mesoscopic Physics, School of Physics, \\Frontiers Science Center for Nano-optoelectronics $\&$ Collaborative Innovation Center of Quantum Matter, Peking University, Beijing 100871, China}

\author{Xinyao Huang}
\email{xinyaohuang@buaa.edu.cn}
\affiliation{School of Physics, Beihang University, Beijing 100191, China}

\author{Kaiye Shi}
\email{sky@pku.edu.cn}
\affiliation{State Key Laboratory of Artificial Microstructure and Mesoscopic Physics, School of Physics, \\Frontiers Science Center for Nano-optoelectronics $\&$ Collaborative Innovation Center of Quantum Matter, Peking University, Beijing 100871, China}
\affiliation{Hefei National Laboratory, Hefei 230088, China}

\author{Qiongyi He}
\email{qiongyihe@pku.edu.cn}
\affiliation{State Key Laboratory of Artificial Microstructure and Mesoscopic Physics, School of Physics, \\Frontiers Science Center for Nano-optoelectronics $\&$ Collaborative Innovation Center of Quantum Matter, Peking University, Beijing 100871, China}
\affiliation{Hefei National Laboratory, Hefei 230088, China}
\affiliation{Collaborative Innovation Center of Extreme Optics, Shanxi University, Taiyuan, Shanxi 030006, China}

\begin{abstract}
Exceptional points (EPs) are non-Hermitian spectral singularities exhibiting fractional-power responses, yet their implications for multipartite entanglement of interacting quantum many-body systems remain largely unexplored.
Here we develop a general framework that links higher-order non-Hermitian degeneracies to the scaling behavior of genuine multipartite entanglement in interacting identical-qubit systems.
Permutation symmetry of the identical qubits decomposes the exponentially large Hilbert space into independent irreducible-representation sectors, thereby constraining the maximal EP order of $N$ qubits to $N+1$ rather than $2^N$.
Near an $n$th-order EP, genuine multipartite entanglement inherits the spectral response and generically exhibits a fractional-power scaling under weak perturbations.
Explicit examples show that conventional two-body interactions support third- and fourth-order EPs with the corresponding entanglement responses, whereas higher-order EPs with genuine multipartite-entangled coalesced states require additional independent interaction channels, such as three-body interactions.
Our results establish a fundamental connection among non-Hermitian degeneracies, multipartite entanglement, and symmetry, extending higher-order EP physics from spectral singularities to genuine many-body quantum correlations.
\end{abstract}

\maketitle
	
\textit{Introduction.---}
Non-Hermitian Hamiltonians provide an effective framework for describing open systems coupled to their environments, as well as postselected quantum dynamics, thereby extending the conventional Hermitian paradigm~\cite{bender_real_1998,ashida_non-hermitian_2020,el-ganainy_non-hermitian_2018}.
One of the clearest departures from Hermitian physics is the appearance of exceptional points (EPs), at which eigenvalues and eigenvectors coalesce and the Hamiltonian becomes defective~\cite{miri_exceptional_2019,ozdemir_paritytime_2019,ding_non-hermitian_2022,bergholtz_exceptional_2021}.
The nonanalyticity of EPs gives rise to branch-point topology and Puiseux spectral responses, leading to a variety of phenomena and applications, including unidirectional mode switching~\cite{doppler_dynamically_2016}, topological state transfer~\cite{yoon_time-asymmetric_2018,xu_topological_2016}, and enhanced sensing~\cite{wiersig_enhancing_2014,chen_exceptional_2017,hodaei_enhanced_2017,wiersig_prospects_2020,lau_fundamental_2018}.
These effects have stimulated extensive studies of EPs in a broad range of classical and engineered platforms, including optical waveguides and photonic structures~\cite{feng_single-mode_2014,hodaei_parity-timesymmetric_2014,peng_loss-induced_2014,miao_orbital_2016,feng_non-hermitian_2017}, acoustic systems~\cite{ding_emergence_2016,tang_exceptional_2020}, electronic circuits~\cite{choi_observation_2018,zhao_exceptional_2024}, and mechanical systems~\cite{cui_experimental_2023}.
Over the past decade, controllable quantum platforms, including superconducting circuits~\cite{naghiloo_quantum_2019,abbasi_topological_2022}, trapped ions~\cite{ding_experimental_2021-1,chen_quantum_2025}, nitrogen-vacancy centers in diamond~\cite{wu_observation_2019,wu_third-order_2024}, and ultracold atoms with engineered dissipation~\cite{li_observation_2019}, have extended non-Hermitian physics into the quantum regime.
Many early quantum demonstrations realized effective two-level non-Hermitian Hamiltonians, whose EPs are restricted to second order~\cite{naghiloo_quantum_2019,ding_experimental_2021-1,wu_observation_2019,li_observation_2019}.
Higher-order EPs, by contrast, exhibit richer spectral topology and stronger Puiseux responses, and have therefore motivated broad theoretical and experimental efforts~\cite{wang_experimental_2023,han_measuring_2024,wu_third-order_2024,chen_quantum_2025,zhang_topological_2025}.

A natural route toward higher-order EPs is to enlarge the effective Hilbert space, either by exploiting multiple internal levels of a single quantum object~\cite{wu_third-order_2024,chen_quantum_2025} or by increasing the number of particles to form a multipartite system~\cite{shi_enhanced_2025,yu_exceptional-point-induced_2026}.
In the latter setting, interaction provides a natural bridge between non-Hermitian singularities and quantum entanglement.
Recent studies~\cite{yu_exceptional-point-induced_2026,zhang_exceptional_2024,shi_quantum_2025,tang_topologically_2024} have revealed a close relation between EP structures and unconventional entanglement phenomena, including maximal bipartite entanglement~\cite{kumar_maximal_2022}, accelerated entanglement generation~\cite{li_speeding_2023,yuan_beating_2026}, and exceptional entanglement transitions~\cite{han_exceptional_2023,yu_exceptional-point-induced_2026,tang_topologically_2024,luo_quantum_2022}.
These investigations, however, have mostly focused on few-body systems with specific two-body interactions, which naturally restricts both the attainable EP order and the scope of quantum correlations, typically to low-order EPs and bipartite or tripartite entanglement.
Going beyond few-body settings, interacting $N$-partite systems naturally call for genuine $N$-partite entanglement, which not only is a key resource for measurement-based quantum computing and quantum-enhanced metrology
~\cite{briegel_measurement-based_2009,giovannetti_quantum-enhanced_2004}, but also provides insight into quantum phase transitions in strongly correlated systems~\cite{deoliveira_multipartite_2006,hauke_measuring_2016}.
This leads to a challenging task: establishing a general framework that connects higher-order non-Hermitian degeneracies with the response of genuine multipartite entanglement (GME).

In this Letter, we provide such a framework for interacting identical-qubit systems.
Permutation symmetry decomposes the Hilbert space into irreducible-representation sectors and bounds the maximal EP order by $N+1$, rather than by the full Hilbert-space dimension $2^N$.
Near an $n$th-order EP (${\rm EP}n$), GME generically inherits the Puiseux-type spectral nonanalyticity and exhibits a fractional-power response to the perturbation strength.
Unlike higher-order EPs assembled from noninteracting qubits, whose coalesced states are separable and carry no GME, the EPs considered here arise in interacting systems and require sufficient tunable interaction channels.
In the collective XYZ model, two-body interactions support ${\rm EP}3$ and ${\rm EP}4$, whereas EPs beyond ${\rm EP}4$ require additional channels such as three-body interactions.
Our results extend the fractional-power response of higher-order EPs from spectral observables to genuine multipartite-entanglement scalings, with potential applications to entanglement-based quantum metrology, sensing, and information processing.

\textit{EPs in interacting multiqubits.---}
Contrary to noninteracting $N$-qubit systems, which can be described solely by single-qubit Hamiltonians~\cite{li_observation_2019,wu_observation_2019,ding_experimental_2021-1,naghiloo_quantum_2019}, the inclusion of interactions necessitates working in the full $2^N$-dimensional Hilbert space. 
While this allows for the emergence of higher-order EPs, it also leads to a rapidly increasing spectral complexity, making the identification of EPs highly nontrivial.
Here we first show that permutation symmetry, which naturally arises in systems of identical elements, provides a powerful physical principle that imposes a strong structural constraint on the Hilbert space~\cite{mandal_symmetry_2021}. This constraint bounds the maximum order of EPs at $N+1$.

\begin{figure}
    \includegraphics[width=1\linewidth]{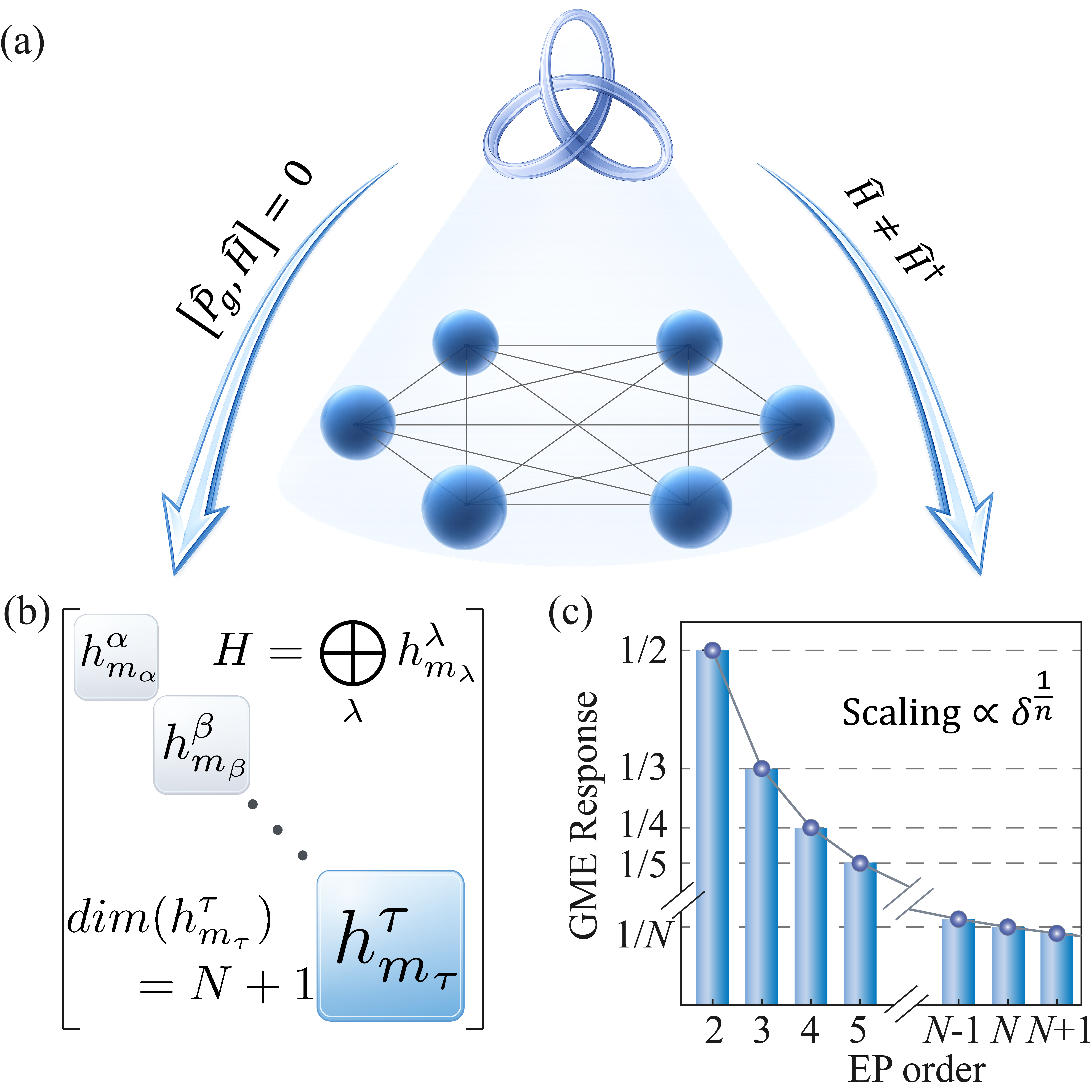}
	\caption{ 
(a) Schematic of interacting identical-qubit systems, where inter-qubit interactions give rise to genuine multipartite entanglement. 
(b) Permutation symmetry gives rise to a universal block structure of the Hamiltonian matrix $H$, decomposing the full Hilbert space into independent symmetry sectors $h_{m_\lambda}^{\lambda}$. The fully symmetric sector $h_{m_\tau}^{\tau}$ corresponds to the largest block with dimension $N+1$.
(c) In the non-Hermitian regime ($\hat H\neq \hat H^{\dagger}$), higher-order EPs can emerge. Near an $n$th-order EP, multipartite entanglement exhibits a fractional-power response to weak perturbations $\propto \delta^{1/n}$, thereby linking the entanglement scaling directly to the EP order.}
	\label{fig1}
\end{figure}

To prove this, we consider a generic non-Hermitian system of $N$ identical qubits with inter-qubit interaction [Fig.~\ref{fig1}(a)]. Here, by an identical-qubit system, we mean a set of equivalent qubits governed by a Hamiltonian invariant under arbitrary qubit permutations. The Hamiltonian is described by ($\hbar=1$)
\begin{equation}
\label{eq:H}
\hat{H}=\sum_{j=1}^{N}\hat{h}_{j}+\hat{H}_{\rm Int},
\end{equation}
where $\hat{h}_{j}$ denotes the single-qubit Hamiltonian. $\hat{H}_{\rm Int}$ accounts for inter-qubit interactions and is
allowed to include arbitrary $l$($\geq 2$)-body interactions~\cite{katz_demonstration_2023,katz_n_2022,menke_demonstration_2022}. 
The non-Hermiticity can be introduced either from $\hat{h}_{j}\neq\hat{h}_{j}^{\dagger}$ or $\hat{H}_{\rm Int}\neq\hat{H}_{\rm Int}^{\dagger}$  as long as the system Hamiltonian [Eq.~\eqref{eq:H}] preserves permutation symmetry, i.e., $\hat{P}_{g}\hat{H}=\hat{H}\hat{P}_{g}, \quad \forall g\in S_{N}$, with $S_{N}$ being the symmetric group and $\hat{P}_{g}$ being the permutation operator.

The key consequence of permutation symmetry is that it imposes a highly constrained structure on the Hamiltonian. Specifically, the Hilbert space of $N$ qubits, ${\cal H}=(\mathbb{C}^2)^{\otimes N}$, naturally carries a unitary representation of the symmetric group $S_N$. 
Since $S_{N}$ is a finite group, this representation is fully reducible and decomposes into a direct sum of irreducible representations (irreps). 
Correspondingly, ${\cal H}$ can be expressed as ${\cal H} = \bigoplus_{\lambda} V_{\lambda}\otimes \mathbb{C}^{m_{\lambda}}$, where $\lambda$ labels the irreps of $S_N$ and is in one-to-one correspondence with integer partitions of $N$, i.e., $\lambda=(p_1,p_2,...,p_k)$ with integers $p_1\geq p_2 \geq ...\geq p_k \geq 0$ and $\sum_{j=1}^k  p_j=N$, $V_{\lambda}$ is the representation space carrying the irrep $\lambda$ of dimension $d_{\lambda}={\rm dim}(V_{\lambda})$, and the non-negative integer $m_{\lambda}$ is the multiplicity with which $V_{\lambda}$ occurs in the decomposition.
According to the Schur-Weyl duality~\cite{hall_review_1940}, the irreps appearing in the decomposition correspond only to partitions with at most two parts, i.e., $\lambda = (N-p,p)$ with $p=0,1,...,\lfloor N/2 \rfloor$. The multiplicity $m_{\lambda}$ of the irrep $\lambda=(N-p,p)$ is $m_{(N-p,p)}=N-2p+1$, yielding the concrete decomposition
\begin{equation}
\label{Hdecom}
    {\cal H} =  \bigoplus^{\lfloor\frac{N}{2}\rfloor}_{p=0} V_{(N-p,p)}\otimes \mathbb{C}^{N-2p+1}.
\end{equation}
This structure imposes strong constraints on any physical observable that respects the permutation symmetry.
In particular, because the system Hamiltonian $\hat{H}$ is invariant under all qubit permutations, Schur’s lemma guarantees that $\hat{H}$ acts as a scalar on each irrep. 
Consequently, the Hamiltonian matrix takes a block-diagonal form [Fig.~\ref{fig1}(b)], with respect to the above decomposition (\ref{Hdecom}): $H =  \bigoplus^{\lfloor{N/2}\rfloor}_{p=0} I_{d_{(N-p,p)}}\otimes h^{(N-p,p)}_{N-2p+1},$ where $I_{d_{\lambda}}$ represents the identity on $V_{\lambda}$, and $h^{\lambda}_{m_{\lambda}}$ is an $m_{\lambda}\times m_{\lambda}$ matrix acting on the multiplicity space $\mathbb{C}^{m_{\lambda}}$ 
(a series of concrete examples can be found in the Supplemental Material~\cite{supplemental}).
That is, permutation symmetry fragments the $2^N$-dimensional Hilbert space into subspaces of dimension at most $N+1$; hence, any Jordan block of the system Hamiltonian $\hat{H}$ has dimension bounded by $N+1$. This yields a universal upper bound on the actual attainable EP order $\le N+1$ for non-Hermitian identical qubits. Higher-order EPs previously discussed in multiqubit systems have mostly relied on noninteracting composite qubits, where $2^N$ eigenvalues coalesce. Such a degeneracy has often been regarded as an EP of order $2^N$, but the actual EP order is set by the largest Jordan block, whose dimension is only $N+1$ for identical qubits. Although interactions are generally expected to lift non-Hermitian degeneracies~\cite{li_speeding_2023,li_multitype_nodate}, we show that the maximal EP order remains $N+1$ with or without interactions.

\textit{Entanglement fractional-power response.---}
The inclusion of interactions allows the eigenstates to exhibit GME, making it natural to examine how such entanglement behaves in the vicinity of an ${\rm EP}n$.
The hallmark of an ${\rm EP}n$ is the $n$th-root topology of the complex-energy Riemann surface, which generically yields a fractional-power eigenvalue splitting under weak perturbations.
Consider a Hamiltonian $\hat{H}_{{\rm EP}n}$ tuned to an ${\rm EP}n$, where $n$ eigenvalues coalesce at $E_0$, while the corresponding eigenvectors coalesce into a single eigenstate $\ket{\psi_0}$, satisfying $\hat{H}_{{\rm EP}n}\ket{\psi_0}=E_0\ket{\psi_0}$.
Introducing a weak perturbation $\delta\hat{O}$ with dimensionless strength $\delta>0$, the Hamiltonian becomes $\hat{H}'=\hat{H}_{{\rm EP}n}+ \delta \hat{O}$, and its eigenvalues admit a fractional-power response~\cite{wiersig_response_2022}, $E_j=E_0 +a_{j}\delta^{\frac{1}{n}}+O(\delta^{\frac{2}{n}})$, with $a_{j}$ is a complex coefficient depending on the eigenvalue branch and the perturbation.

We now connect this spectral singularity to the GME degree of the eigenstate.
Substituting the perturbed energy into the eigenvalue equation, $(\hat{H}_{{\rm EP}n}+\delta\hat{O})\ket{\psi_j}=E_j\ket{\psi_j}$, we obtain the corresponding eigenstate for $E_j$ as $\ket{\psi_j}=\ket{\psi_0} +\delta^{1/n}\ket{\psi_{1,j}}+O(\delta^{2/n})$, where $\ket{\psi_{1,j}}$ is a Jordan vector~\cite{wiersig_response_2022}, which satisfies $(\hat{H}_{{\rm EP}n}-E_0\hat{I})\ket{\psi_{1,j}}=a_{j}\ket{\psi_0}$ with the identity operator $\hat{I}$.
Retaining the dominant term of the perturbation $\delta^{1/n}$, the density matrix is $\rho_j=(\rho_0+\delta^{1/n}\rho_{1,j})/{\cal N}_{\delta,j}$, 
with $\rho_0=\ket{\psi_0}\bra{\psi_0}$, $\rho_{1,j}=\ket{\psi_0}\bra{\psi_{1,j}}+\ket{\psi_{1,j}}\bra{\psi_0}$, and ${{\cal N}_{\delta,j}}={\rm Tr}(\rho_0+\delta^{1/n}\rho_{1,j})$. 
Since the eigenstate can be regarded as an $N$-partite pure state, i.e., $\ket{\psi_j}\in  {\cal H}_1\otimes...\otimes{\cal H}_N$ with $m$-th qubit subspace ${\cal H}_m$, its GME degree 
can be quantified by the genuine multipartite concurrence, defined as~\cite{ma_measure_2011} ${\cal C}(\ket{\psi_j}) := \min_{A|\bar A} \sqrt{2\bigl[1 - \operatorname{Tr}_{A}(\rho_{A,j}^2)\bigr]}$, where the minimum is evaluated across all bipartitions $A|\bar A$ of the $N$ parties. 
An arbitrary bipartition $A|\bar A$ splits the $N$ parties into two groups, with associated Hilbert spaces ${\cal H}_A=\bigotimes_{j\in A}{\cal H}_{j}$ and ${\cal H}_{\bar A}=\bigotimes_{l\in {\bar A}}{\cal H}_{l}$, satisfying ${\cal H}={\cal H}_A\otimes{\cal H}_{\bar A}$, and $\rho_A$, the reduced density matrix of subsystem $A$, is defined as the partial trace of the full density matrix over $\bar A$, i.e., $\rho_{A,j} = {\rm Tr}_{\bar A}(\rho_{j})\nonumber=(\rho_{A_0,j}+\delta^{1/n}\rho_{A_1,j})/{\cal N}_{\delta,j}$, with $\rho_{A_0,j}={\rm Tr}_{\bar A}(\rho_0)$ and $\rho_{A_1,j}={\rm Tr}_{\bar A}(\rho_{1,j})$. 
We consider the response of GME degree near an ${\rm EP}n$, defined as $\Delta {\cal C}(\ket{\psi_j}) = {\cal C}(\ket{\psi_j}) - {\cal C}(\ket{\psi_0})$. 
Denote by $A^*|\bar A^*$ the bipartition that minimizes $\sqrt{2[1-{\rm Tr}_{ A}(\rho_{A_0,j}^2)]}$. 
For sufficiently small $\delta$, and assuming that this minimizing bipartition remains unchanged, the GME degree of the eigenstate can be expressed as ${\cal C}(\ket{\psi_j})=\sqrt{2[1-{\rm Tr}_{A^*}(\rho_{A^*,j}^2)]}$. 
If the coalesced EP eigenstate has nonzero GME, ${\rm Tr}_{A^*}(\rho^2_{A_{0}^*,j})\neq 1$, so the square-root expression is regular.
The leading order in the perturbation can be derived as 
\begin{equation}
\Delta {\cal C}(\ket{\psi_j}) =k_{1,j} \delta^{\frac{1}{n}} + O(\delta^{\frac{2}{n}}),
\end{equation}
where $k_{1,j}$ is a branch-dependent coefficient given in the Supplemental Material~\cite{supplemental}.


In summary, the entanglement of eigenstates near an EP generically follows a fractional-power law, with leading dependence determined by the EP order $n$---a direct manifestation of the singularity's mathematical structure. Whereas a diabolic point (DP) in a Hermitian system permits an ordinary Taylor expansion, $\sum_{l=1}^{\infty}k_l\delta^l$, the nonanalyticity at an $n$th‑order EP forces the expansion to take the form of a Puiseux series, $\sum_{l=1}^{\infty}k_l\delta^{l/n}$.


\textit{Examples.---}We now proceed to illustrate our theory with specific examples, exhibiting higher-order EPs and the associated GME response.
Although we have previously argued that the symmetry-allowed upper bound on the attainable EP order for $N$ identical qubits is $(N+1)$, it does not guarantee that such degeneracies can be achieved with arbitrary interactions.
Therefore, we first identify the conditions required for the emergence of an EP of order $N+1$. We focus on the fully symmetric subspace [i.e., $\lambda=(N)$] and consider the Hamiltonian $\hat{h}^{(N)}_{N+1}$ acting within it, whose characteristic polynomial is $\sum_{k=0}^{N+1} c_k E^{N+1-k}=0$, where the coefficients $c_k$ depend explicitly on the specific form of the corresponding matrix $ h^{(N)}_{N+1}$.
An $(N+1)$th-order EP demands that $(N+1)$ eigenvalues degenerate and the dimension of the corresponding eigenspace be one. This implies that the characteristic polynomial has a root $\varepsilon$ of algebraic multiplicity $(N+1)$ and geometric multiplicity (corresponding to the number of Jordan blocks) equal to one. 
Applying Vi\`{e}te's formula, $E_1+...+E_{N+1}=-c_1/c_0$, the $(N+1)$-fold degenerate eigenvalue is $\varepsilon= -c_1/[(N+1)c_0]$, whose geometric multiplicity is given by $m_g(\varepsilon)=N+1- {\rm rank}[\varepsilon I_{N+1}-h^{(N)}_{N+1}]$.
Thus, the appearance of an $(N+1)$th-order EP requires

\begin{equation}
\begin{aligned}
\label{AM}
\frac{(N+1)!}{k!(N+1-k)!}
\left[\frac{c_1}{(N+1)c_0}\right]^k
&= \frac{c_k}{c_0}, \quad k=2,\ldots,N+1,\\
{\rm rank}\left[-\frac{c_1 I_{N+1}}{(N+1)c_0}
-h^{(N)}_{N+1}\right]&=N .
\end{aligned}
\end{equation}

\begin{figure}[!t]
    \centering
    \includegraphics[width=\linewidth]{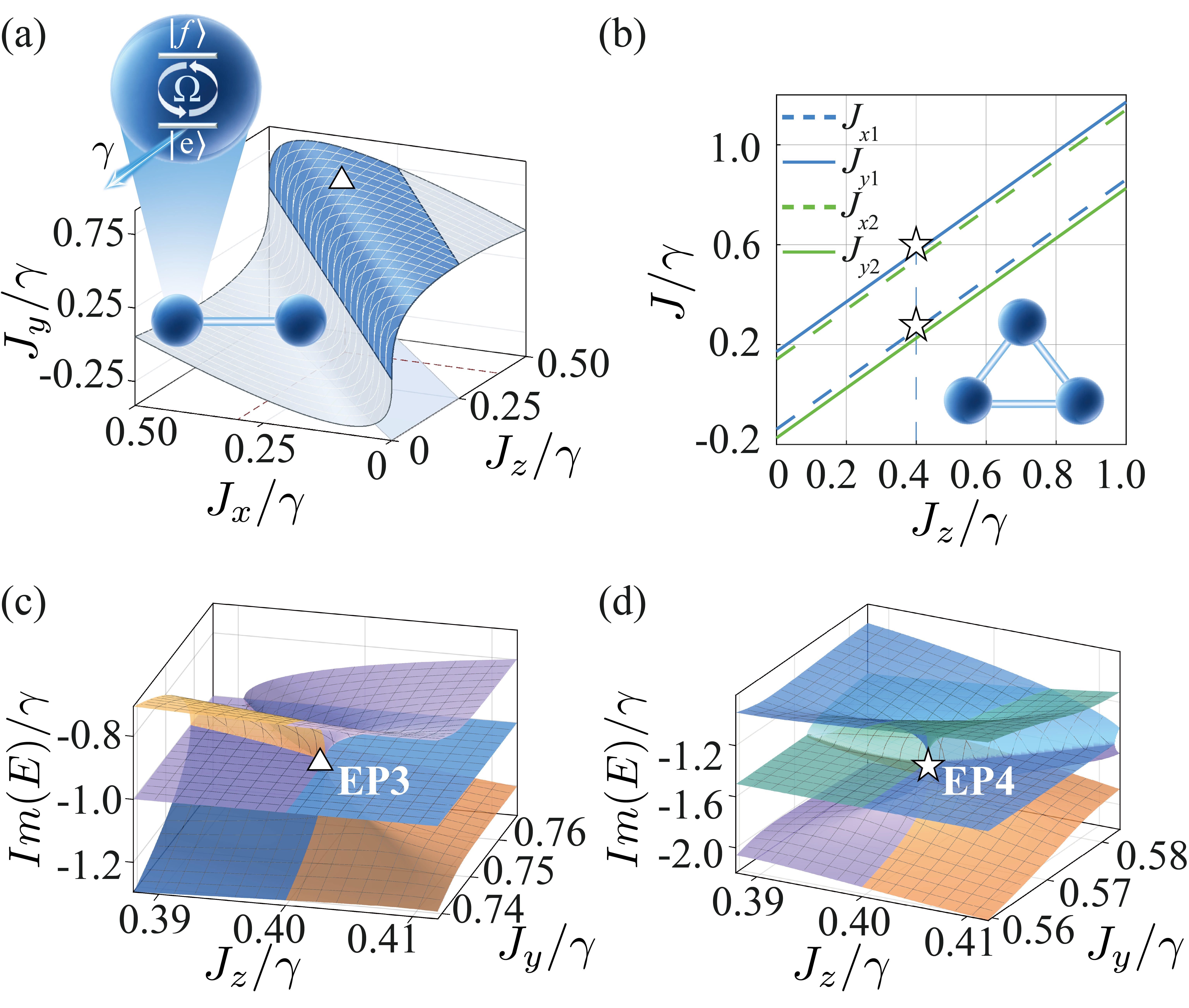}
    \caption{
    (a) Schematic of an interacting two-qubit system and the corresponding EP conditions projected onto the interaction-parameter subspace $(J_x,J_y,J_z)$, with $\gamma$ as the energy unit. The colored surface denotes the manifold of ${\rm EP}3$s, with the blue region illustrating the real-$\Omega$ solutions selected from Eq.~\eqref{AM}.
    (b) EP conditions for the interacting three-qubit system in the same parameter subspace. Two exceptional lines of ${\rm EP}4$s are obtained by expressing $(J_x,J_y)$ as functions of $J_z$. The white triangle in (a) and the stars in (b) mark the parameters of the ${\rm EP}3$ and ${\rm EP}4$ used in (c) and (d), respectively.
    (c),(d) Imaginary parts of the n-sheeted Riemann surfaces of the complex eigenvalues in the vicinity of the marked ${\rm EP}3$ and ${\rm EP}4$ with $(J_x,J_z)/\gamma=(0.3,0.4)$ and $J_z/\gamma=0.4$ fixed respectively, showing the characteristic three- and four-sheet structures.
    The real parts of the Riemann surfaces are shown in Fig.~S1 of the Supplemental Material~\cite{supplemental}.
    }
    \label{fig2}
\end{figure}

The conditions derived above, however, are not always satisfiable in the interacting multiqubit systems described by Eq.~\eqref{eq:H}, as will become clear in the examples that follow.
We focus on a paradigmatic non-Hermitian model directly inspired by experiments.
The single-qubit part reads $\hat{h}_j=\Omega \hat{\sigma}_j^x -i\gamma \hat{\sigma}^-_j\hat{\sigma}_j^+$~\cite{ding_experimental_2021-1,wu_observation_2019,naghiloo_quantum_2019}, with 
$\hat{\sigma}^x_j=\hat{\sigma}^+_j+\hat{\sigma}^-_j$ and $\hat{\sigma}^{+(-)}_j= \ket{f(e)}_j\bra{e(f)}$. Here $\ket{e(f)}_j$ form an orthonormal basis spanning the Hilbert space of the $j$th qubit; they are resonantly coupled by a driving field of strength $\Omega$ and $\gamma$ is the dissipation rate of the state $\ket{e}$, which is set as the energy unit [Fig.~\ref{fig2}(a)].
The inter-qubit interaction is taken as the generic XYZ Heisenberg interaction that appears widely in quantum simulation platforms~\cite{luo_hamiltonian_2025,monroe_programmable_2021,georgescu_quantum_2014,altman_quantum_2021}, i.e., $\hat{H}_{\rm Int} = \hat{H}_{\rm XYZ}=-\sum_{j\neq l,\alpha =x,y,z}J_{\alpha} \hat{\sigma}^{\alpha}_j \hat{\sigma}^{\alpha}_l$.
For $N=2$, Eq.~\eqref{AM} reduce to two independent constraints, defining a two-dimensional ${\rm EP}3$ surface in the four-dimensional parameter space $(\Omega,J_x,J_y,J_z)$~\cite{supplemental}.
As illustrated in Fig.~\ref{fig2}(a), we project this surface onto the $(J_x,J_y,J_z)$ subspace, where the blue region denotes the parameter regime in which the corresponding solutions for $\Omega$ are real. 
Extending to interacting three-qubit systems, Eq.~\eqref{AM} provide three independent equations~\cite{supplemental}. Their solutions yield $|\Omega|/\gamma = 1 - \sqrt{3}/2$, which is independent of the interaction parameters $(J_x, J_y, J_z)$. 
As demonstrated in Fig.~\ref{fig2}(b), two distinct exceptional lines appear in the interaction parameter space, and each point on these lines corresponds to an ${\rm EP}4$. 
Figures~\ref{fig2}(c) and \ref{fig2}(d) show the imaginary parts of the complex-eigenvalue Riemann surfaces near the marked EP$3$ and EP$4$, revealing the characteristic cubic- and fourth-root branch-point structures, respectively.



\begin{figure}
\includegraphics[width=1\linewidth]{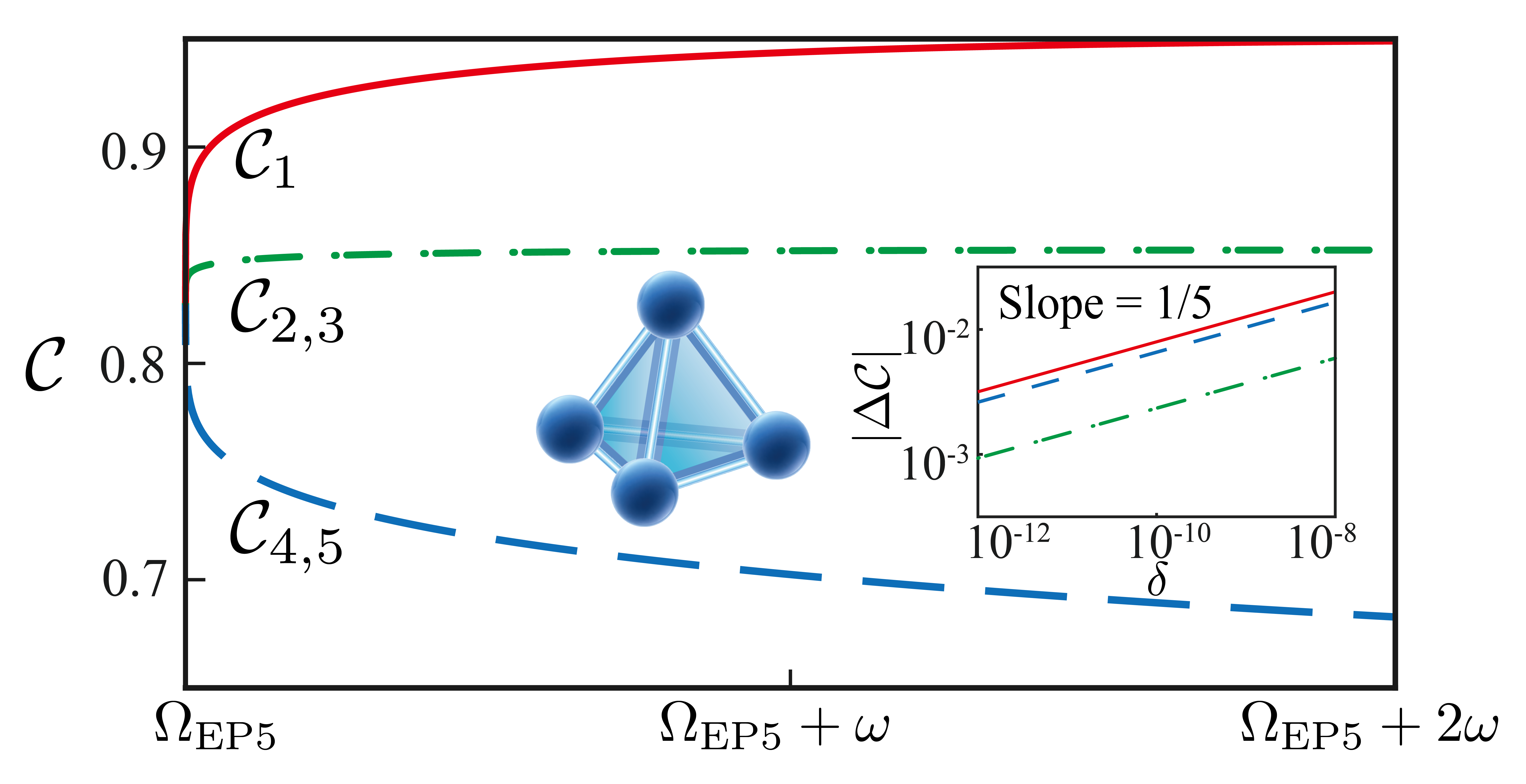}
 \caption{Schematic of the interacting four-qubit system with three-body interactions and its genuine multipartite concurrence response of the five eigenstates, ${\cal C}_i$ ($i=1,\dots,5$), near an ${\rm EP}5$ with $\omega/\gamma=2 \times10^{-4}$. 
The inset shows the corresponding log-log behavior, confirming the fractional-power scaling $|\Delta {\cal C}| \propto \delta^{1/5}$ with the perturbation strength $\delta=(\Omega-\Omega_{{\rm EP}5})/\gamma$. }
\label{fig3}
\end{figure}

However, for $N>3$, Eq.~\eqref{AM} admit only a trivial isotropic solution, whose coalesced eigenstate is separable (see the Supplemental Material~\cite{supplemental}).
Consequently, for a $(N+1)$th-order EP with $N\ge 4$, GME in the eigenstate requires interaction channels beyond two-body terms.
This restriction follows from the fact that the number of independent algebraic constraints in Eq.~\eqref{AM} grows with $N$ and eventually surpasses the number of independently tunable parameters available in $\hat{H}$.
For $N=4$, for instance, the XYZ interaction alone cannot supply enough degrees of freedom to realize ${\rm EP}5$, so that additional multibody interactions become indispensable.
A natural candidate is the three-body term $\hat{H}_{\mathrm{3-body}} = -\eta \sum_{j\neq k,\;k\neq l,\;l\neq j} \hat{\sigma}_j^{x}\hat{\sigma}_k^{x}\hat{\sigma}_l^{x}$, which has already been demonstrated experimentally in trapped-ion platforms~\cite{katz_demonstration_2023,katz_n_2022}.
The extended interaction Hamiltonian $\hat{H}_{\mathrm{Int}} = \hat{H}_{\mathrm{XYZ}} + \hat{H}_{\mathrm{3-body}}$ provides the necessary additional control to satisfy the EP conditions and to realize ${\rm EP}5$ with GME. 
We further confirm the entanglement scaling law by perturbing the ${\rm EP}5$ realized in a four-qubit system with Hamiltonian $\hat{H}' = \hat{H}_{\mathrm{EP}5} + \delta \hat{O}$, where $\hat{O} = \gamma \sum_{j} \hat{\sigma}_{j}^{x}$ and $\delta = (\Omega - \Omega_{\mathrm{EP}5})/\gamma$.
Figure~\ref{fig3} shows that the GME $\mathcal{C}_i$ ($i = 1,\dots,5$) responds sharply near the ${\rm EP}5$.
By considering the logarithmic behavior of this response (see the inset of Fig.~\ref{fig3}), we find that the slopes of these curves are all equal to $1/5$, thus confirming our derivation of the fractional-power response for GME near an EP$n$ (see the Supplemental Material for more examples~\cite{supplemental}).



\textit{Concluding remarks.---}  
We have developed a general framework connecting higher-order EPs with GME responses in non-Hermitian interacting identical-qubit systems. 
Permutation symmetry bounds the largest Jordan block by $N+1$, while an ${\rm EP}n$ produces the fractional entanglement response $\Delta {\cal C}\propto \delta^{1/n}$. 
Our all-to-all collective qubit models verify these predictions and show that high-order EPs with genuinely multipartite-entangled coalesced eigenstates require sufficiently many independently tunable interaction channels, which at higher orders go beyond conventional two-body interactions
Beyond the general theoretical framework, the interacting qubit models considered here are closely connected to existing state-of-the-art quantum platforms.
Cavity-mediated platforms naturally realize collective all-to-all spin couplings~\cite{Norcia2018,Sauerwein2023}, and have recently enabled XYZ two-body interactions~\cite{luo_hamiltonian_2025}, as well as engineered three- and four-body interactions~\cite{luo_realization_2025}.
Trapped-ion platforms provide another promising route, where effective spin-spin interactions can be tuned over a long range and approximately scale as $1/r^\alpha$, with the limit $\alpha\rightarrow0$ approaching all-to-all coupling~\cite{monroe_programmable_2021}. Moreover, both non-Hermitian single-spin Hamiltonians and engineered multibody spin interactions have already been experimentally realized in trapped-ion systems~\cite{ding_experimental_2021-1,katz_demonstration_2023,katz_n_2022}. 
These experimental advances provide promising routes toward observing fractional entanglement scaling near higher-order EPs. More broadly, our work bridges non-Hermitian spectral physics and multipartite entanglement as a functional resource, opening new directions for quantum sensing, metrology, and information processing.

\indent 

\textit{Acknowledgement.---}
We acknowledge helpful discussions with Shuming Cheng.
This work was supported by Beijing Natural Science Foundation (Grant No.~Z240007), National Natural Science Foundation of China (No.~12125402, No.~12534016, No.~12504585, No.~12474354), Quantum Science and Technology-National Science and Technology Major Project (Grant No.~2024ZD0302401 and No.~2021ZD0301500), and the Fundamental Research Funds for the Central Universities. K.S. acknowledges support from the China Postdoctoral Science Foundation (Grant No. 2025M783357).
\bibliography{ref}


\end{document}